\documentclass[11pt]{article}
\begin{document}

\title{\textbf{\textsf{Comments on ``Charged particle dynamics in the field of slowly rotating compact star"}}}
\author{ Mubasher Jamil\footnote{mjamil@camp.edu.pk} and Asghar
Qadir\footnote{aqadirmath@yahoo.com}
\\ \\
\textit{\small Center for Advanced Mathematics and Physics}\\
\textit{\small National University of Sciences and Technology}\\
\textit{\small Peshawar Road, Rawalpindi, 46000, Pakistan} \\
} \maketitle
\begin{abstract}
B.M. Mirza \cite{mirza} presented a solution of coupled
Einstein-Maxwell equations for a slowly rotating neutron star;
however his derivations had some errors and implicit assumptions
that rendered the solution invalid. We point out the errors and
present a mathematically consistent solution. The resulting solution
is also physically consistent as it remains finite in the no
rotation limit, whereas Mirza's solution diverges for zero rotation.
\end{abstract}

Slowly rotating neutron stars were first investigated in\ 1967 as a
slow rotation approximation $(0<$ $a<1,O(a^{2}))$ to the Kerr metric
\cite{hartle} and have astrophysical relevance as most of the
observed pulsars are actually slowly rotating relative to the speed
of light (one fifth for a millisecond pulsar) \cite{rees}. Charged
particle dynamics around such stars is investigated by constructing
the corresponding Einstein-Maxwell equations assuming the slow
rotation approximation \cite{ress}. Later Mirza \cite{mirza} solved
the same model using an ansatz. The chosen ansatz was dimensionally
inconsistent and yielded an unphysical answer for the fields which
gives a divergent expression for the radiation emitted in the no
rotation limit. We prove that his chosen ansatz, even after
dimensional modification, gives a divergent result in the no
rotation limit. We then suggest an ansatz that avoids divergences in
any limit and is physically more\ meaningful. \pagebreak

 The
spacetime exterior to a slowly rotating neutron star is given by the
slow rotation approximation to the Kerr metric:

\begin{equation}
ds^{2}=-e^{2\Phi (r)}dt^{2}-2\omega (r)r^{2}\sin ^{2}\theta
dtd\varphi +r^{2}\sin ^{2}\theta d\varphi ^{2}+e^{-2\Phi
(r)}dr^{2}+r^{2}d\theta ^{2}, \label{1}
\end{equation}%
where

\begin{equation}
e^{2\Phi (r)}=(1-\frac{2M}{r}),  \label{2}
\end{equation}%
and
\begin{equation}
\omega (r)\equiv \frac{d\varphi }{dt}=-\frac{g_{t\varphi
}}{g_{\varphi \varphi }},  \label{3}
\end{equation}%
is the angular speed of a freely falling frame brought into rotation
by
frame dragging. Here and in what follows the Greek indices run as $%
t,r,\theta ,$ and $\varphi $ respectively and we use gravitational
units in which $G=1=c$. The general relativistic form of the Maxwell
equations is:
\begin{equation}
F_{\alpha \beta ,\gamma }+F_{\beta \gamma ,\alpha }+F_{\gamma \alpha
,\beta }=0,  \label{4}
\end{equation}

\begin{equation}
\left( \sqrt{-g}F^{\alpha \beta }\right) _{,\beta }=4\pi
\sqrt{-g}J^{\alpha },  \label{5}
\end{equation}%
where $g$ is the determinant of the metric tensor $g_{\alpha \beta
}$ given by the Einstein equations and $J^{\alpha }$ is 4-vector
current density. Here $F_{\alpha \beta }$ is the generalized
electromagnetic field tensor for an ideal fluid given by a unique
tensorial expression:

\begin{equation}
F_{\alpha \beta }=u_{\alpha }E_{\beta }-u_{\beta }E_{\alpha }+\eta
_{\alpha \beta \gamma \delta }u^{\gamma }B^{\delta },  \label{6}
\end{equation}%
where $\eta _{\alpha \beta \gamma \delta }$ is the totally skew
tensor,
 $%
E_{\alpha }$ and $B^{\alpha }$ are the electric and magnetic fields
and
 $%
u^{\alpha }$ is the unit velocity 4-vector \cite{landau}. In general
$J^{\alpha }$ is the sum of two terms corresponding to a convection
and a conduction current:

\begin{equation}
J^{\alpha }=\epsilon u^{\alpha }+\sigma u_{\beta }F^{\beta \alpha }
\label{7}
\end{equation}%
where $\epsilon $ is the proper charge density, $\sigma $ is the
conductivity of the fluid. For a zero angular momentum observer
(ZAMO)
 $%
u_{r} $ and $u_{\theta}$ vanish, and using $u^{\alpha }u_{\alpha
}=-1$, the components of the 4-velocity vector are:
\begin{equation}
u^{\alpha }=e^{-\Phi (r)}(1,0,0,\omega (r)),\quad u_{\alpha
}=e^{\Phi (r)}(-1,0,0,0).  \label{8}
\end{equation}%
The electromagnetic field outside the neutron star is now determined
by eqs.(4) and (5). To solve this system of equations let us assume
Mirza's ansatz
modified to maintain dimensional consistency, for the electric field $%
\mathbf{E}$:

\begin{equation}
E_{r}(r,\theta )\equiv k_{1}E_{\theta }(r,\theta )\equiv
k_{2}E_{\varphi }(r,\theta )=R_{E}(r)\Theta _{E}(\theta ), \label{9}
\end{equation}%
where $k_{1}$ and $k_{2}$ are some dimensional constants. For the
magnetic field $\mathbf{B}$, the corresponding\ ansatz will be
\begin{equation}
B_{r}(r,\theta )\equiv k_{3}B_{\theta }(r,\theta )\equiv
k_{4}B_{\varphi }(r,\theta )=R_{B}(r)\Theta _{B}(\theta ),
\label{10}
\end{equation}%
where $k_{3}$ and $k_{4}$ are some dimensional constants. Further,
following Mirza, we take a constant angular speed of rotation,
$\omega _{\circ }$.

Solving eqs.(4) and (5) using eqs.(9) and (10), we get

\begin{equation}
E_{r}\equiv k_{1}E_{\theta }\equiv k_{2}E_{\varphi }=R_{E}(r)\Theta
_{E}(\theta )=\frac{A_{1}}{\omega _{\circ }r^{2}\sin \theta },
\label{11}
\end{equation}%
\
\begin{equation}
B_{r}\equiv k_{3}B_{\theta }\equiv k_{4}B_{\varphi }=R_{B}(r)\Theta
_{B}(\theta )=\frac{A_{2}}{\omega _{\circ }r^{2}\sin \theta },
\label{12}
\end{equation}%
where $A_{1}\ $and $A_{2}$ are arbitrary constants.\ It can easily
be seen that in the no rotation limit i.e $\omega _{\circ
}\rightarrow 0$, the electric and magnetic fields become infinite,
which cannot be true, because it yields infinite energy radiated by
a non-rotating neutron star. Further, the Poynting vector
$\mathbf{S}=\mathbf{E}\times \mathbf{B}$ \ gives the momentum flux
and hence the energy radiated $\varepsilon $. It yields
 $%
\varepsilon \propto \frac{1}{\omega _{\circ }^{2}}$\, where the
proportionality factor depends upon $k_{1}...k_{4}$ and $(r,\theta
).$ If it is non zero, the radiated energy diverges as $\omega
_{\circ
 }\rightarrow 0$%
\ which is impossible, and if $\varepsilon =0,$ the rotating star
would not radiate and hence would not be a model for a pulsar.\
Infact, in general, for a rotating neutron star $\frac{{d\varepsilon
}}{{dt}}\propto \omega ^{6}$ \cite{rees}. Hence Mirza's ansatz does
not work even after correcting for dimensions.

We present another ansatz that avoids the above mentioned
impossibilities.
\begin{equation}
\mathbf{E}=(E_{r},0,E_{\varphi }),\mathbf{B}=(0,B_{\theta },0),
\label{13}
\end{equation}%
where $\mathbf{E}$ and $\mathbf{B}$\ has $r$ and $\theta $
dependence only. To solve eqs.(4) and (5) we shall use the following
separation\ ansatz for electric and magnetic fields

\begin{equation}
E_{r}(r,\theta )\equiv k_{5}E_{\varphi }(r,\theta )=R_{E}(r)\Theta
_{E}(\theta ),  \label{14}
\end{equation}%
\begin{equation}
B_{\theta }(r,\theta )=R_{B}(r)\Theta _{B}(\theta ).  \label{15}
\end{equation}%
Solving eqs.(4) and (5) using eqs.(13), (14) and (15) we get

\begin{equation}
E_{r}\equiv k_{5}E_{\varphi }=R_{E}(r)\Theta _{E}(\theta
 )=\frac{A_{3}}{%
r^{2}u^{t}},  \label{16}
\end{equation}

\begin{equation}
B_{\theta }=R_{B}(r)\Theta _{B}(\theta )=\frac{A_{4}\omega _{\circ }}{%
u_{t}\sin \theta },  \label{17}
\end{equation}%
where $A_{3}\ $and $A_{4}$ are arbitrary constants. Hence,
$\mathbf{E}$
 and $%
\mathbf{B}$ remain finite in the no-rotation limit and the radiation
vanishes for a non-rotating object, as required.

There will be several other ansatz that would yield solutions to the
system of equations but care must be taken to avoid non-physical
solutions such as Mirza obtained. As an extension to the problem,
one can solve the above system using the source terms i.e.\
$j^{\alpha }\neq 0$ and deduce some physically interesting
realizable results.
\section*{Acknowledgments} One of us (MJ) is grateful to Prof. Muneer
A. Rashid and Ibrar Hussain for enlightening discussions.

\end{document}